\documentclass[showpacs,preprintnumbers,amsmath,amssymb,twocolumn,superscriptaddress,pra]{revtex4}
\usepackage{graphicx}
\usepackage{amsfonts}
\usepackage{amsmath, amsthm, amssymb}
\usepackage{dsfont}
\usepackage{subfigure}

\newcommand{\vk}{{\boldsymbol{k}}} 
\newcommand{\vecr}{\boldsymbol{r}}

\newcommand{\vq}{\boldsymbol{q}} 
\newcommand{\e}[1]{\mathrm{e}^{#1}}

\newcommand{\eg}{\textit{e.g. }}

\def\i{\mathrm{i}}

\begin{document}
\title[Probing phase-separation in Bose-Fermi mixtures by the critical superfluid velocity]
{Probing phase-separation in Bose-Fermi mixtures by the critical superfluid velocity}
\author{Jacob Linder}
\affiliation{Department of Physics, Norwegian University of
Science and Technology, N-7491 Trondheim, Norway}
\author{Asle Sudb{\o}}
\affiliation{Department of Physics, Norwegian University of
Science and Technology, N-7491 Trondheim, Norway}

\date{Received \today}
\begin{abstract}
We investigate the effect exerted by spin-polarized fermions on the interaction between superfluid bosons for a Bose-Fermi mixture residing on an optical lattice, with particular emphasis on the possibility of an induced phase-separation. Using a set of microscopic parameters relevant to a $^{40}$K-$^{87}$Rb mixture, we show how the phase-separation criterion may be directly probed by means of the critical superfluid velocity of the 
bosonic condensate. We report quantitative results for the magnitude of the superfluid velocity and its dependence on the trap depth, the  
boson-fermion interaction, and the fermionic filling fraction. All of these parameters can be controlled experimentally in a well-defined 
manner. We propose an experimental setup for probing the critical superfluid velocity.
\end{abstract}
\pacs{03.75.Lm, 05.30.Jp, 05.30.Fk}

\maketitle

\section{Introduction}
The scenario of trapped ultracold atoms residing on tunable optical lattices offers a fertile arena for exploration of fundamental physics. One of the most intriguing features of such systems is the possibility to exert experimental control over the environment where the atoms are located. This is accomplished by means of tuning the frequency of the lasers generating the optical lattice, a feature which may be used to induce phase transitions in the system. Trapped ultracold atoms host phases including supersolidity, Mott insulation, and superfluidity, and have been studied extensively (see Refs. \cite{morsch_rmp_06, bloch_rmp_08, giorgini_rmp_08} and references therein).
\par
The atoms on the optical lattice may be either bosons, fermions, or a mixture of both. In particular, it is experimentally possible to generate Bose-Fermi mixtures where the spins of the fermions are frozen due to the influence of the confining magnetic trap. Depending on the relative strength of the intersite hopping and interaction parameters, respectively, the system enters into a specific quantum phase. A key observation in this context is that the interaction between the fermions and the bosons may strongly influence the preferred ground-state of the system. 
\par
A convenient way of treating Bose-Fermi mixtures theoretically is to integrate out the fermionic sector, thus obtaining an effective interacting boson theory. This is possible when the fermion-spin is frozen, leading to a vanishing fermionic on-site interaction. It turns out that the resulting effective boson-boson interaction $U_{b}$ is very sensitive to the presence of a fermionic density \cite{buchler_prl_03}. In fact, the boson-fermion interaction may render the Bose system thermodynamically unstable and lead to phase-separation provided the effective boson-boson interaction becomes attractive. Such a phase-separation is certainly interesting in its own right, but also represents a serious obstacle for observing novel quantum phases arising out of an interacting mixture of bosons and fermions, since it narrows the parameter range in which the bosons and fermions coexist.
\par
From an experimental point of view, the phase-separation criterion may be probed by means of monitoring the critical superfluid velocity in the Bose-Fermi mixture. This has previously been accomplished experimentally in Bose-Einstein condensates by means of stirring the trapped gas with a blue-tuned laser \cite{raman_prl_99, onofrio_prl_00}. The superfluid quantum state then becomes energetically unstable at a critical magnitude of 
the velocity. Previously, several aspects of the critical superfluid velocity have been investigated in the context of single- and multicomponent Bose-Einstein condensates \cite{menotti_pra_04, boers_epl_04, kaurov_prl_05, fil_pra_05, ruostekoski_pra_07, liu_pra_07, kravchenko_jltp_08, linder_pra_09, shrestha_arxiv_09} as well as in Fermi superfluids \cite{ghosh_pra_06, diener_pra_08, silva_pra_09}. 
\par
However, an analysis of the critical velocity for the bosonic superfluid phase in a Bose-Fermi mixture is still lacking. Of particular interest 
is the question of how the fermion-boson interaction influences the critical velocity in a Bose-Fermi mixture. Very recently, it was shown in Ref. \cite{best_prl_09} how the fermion-boson interaction can be tuned over a wide range using a Feshbach resonance, allowing for both an attractive 
or repulsive character. This finding opens up new possibilities in terms of probing the various quantum phases that may arise in such 
Bose-Fermi mixtures \cite{sengupta_pra_07}.
\par
In this paper, we calculate quantitatively the critical superfluid velocity $v_c$ in a Bose-Fermi mixture using a set of realistic experimental parameters pertaining to a $^{40}$K-$^{87}$Rb mixture. We focus especially on how $v_c$ depends on the trap depth, the boson-fermion interaction, 
and the fermionic filling fraction, which all are parameters that can be tuned experimentally in a controllable fashion. Our results yield 
numbers which are similar in magnitude to the critical velocity obtained experimentally in a Bose-Einstein condensate \cite{raman_prl_99}, 
namely of order $\mathcal{O}$(mm/s). We also propose an experimental setup for probing the critical superfluid velocity in a Bose-Fermi 
mixture.
\par
This work is organized as follows. In Sec. \ref{sec:theory}, we introduce the theoretical framework and previously obtained results which we will rely on in our study of the critical superfluid velocity. In Sec. \ref{sec:results}, we present our main results, which is a study of how $v_c$ is influenced by the trap depth, the boson-fermion interaction, and the fermionic chemical potential. We discuss our results in Sec. \ref{sec:discussion}, suggesting also a possible experimental setup which may probe the predicted effects, and finally conclude in Sec. \ref{sec:summary}. Note that in order to obtain quantitative results for $v_c$, we will not use units such that $\hbar=c=1$, but instead use their actual values.

\begin{figure}[t!]
\centering
\resizebox{0.5\textwidth}{!}{
\includegraphics{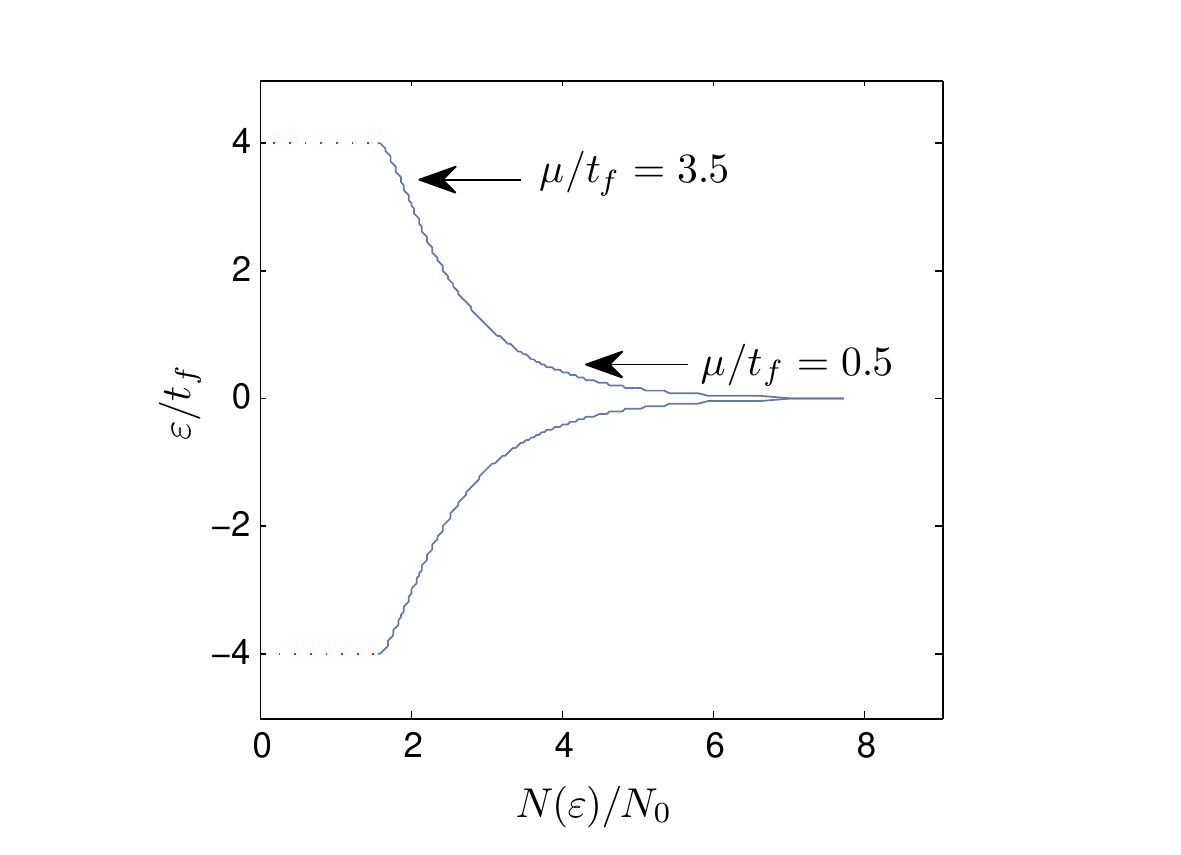}}
\caption{(Color online) Fermionic density of states on a square lattice, featuring a van Hove singularity at $\varepsilon=0$. We consider two filling fractions characterized by $\mu/t_f=0.5$ and $\mu/t_f=3.5$, respectively. Here, $N_0=1/(2\pi^2t_f)$. }
\label{fig:DOS}
\end{figure}

\section{Theory}\label{sec:theory}

To begin with, we briefly account for the route employed to obtain our main results. A general Hamiltonian describing interacting fermions and bosons reads 
\begin{align}
H = H_f + H_b + H_{bf},
\end{align}
where we have defined
\begin{align}
H_f = \int \text{d}\vecr \psi^\dag_f(\vecr)[-\hbar^2\nabla^2/(2m_f)+ V_f(\vecr)]\psi_f(\vecr),
\end{align}
and $f\to b$ for $H_b$. The interaction term is 
\begin{align}\label{eq:Hint}
H_{bf} &= \int \text{d}\vecr \Big[g_{bf}\psi_b^\dag(\vecr)\psi_b(\vecr)\psi^\dag_f(\vecr)\psi_f(\vecr)\notag\\
&+ g_{b}\psi^\dag_b(\vecr)\psi^\dag_b(\vecr)\psi_b(\vecr)\psi_b(\vecr)\Big].
\end{align}
Above, $m_\alpha$ and $V_\alpha$ denote the mass and optical lattice potential for $\alpha=\{f,b\}$, whereas $g_{b}$ and $g_{bf}$ are the 
boson-boson and boson-fermion interactions, respectively. It is implicitly assumed above that we are dealing with a fully spin-polarized 
fermion system. By expanding the field operators $\psi_\alpha$ in Bloch wavefunctions $\{u_\vk,v_\vk\}$ for a periodic potential 
\cite{buchler_pra_04},
\begin{align}
\psi_f^\dag(\vecr) &= \frac{1}{\sqrt{N}}\sum_\vk v_\vk(\vecr)c_\vk^\dag,\notag\\
\psi_b^\dag(\vecr) &= \frac{1}{\sqrt{N}}\sum_\vk u_\vk(\vecr)b_\vk^\dag,
\end{align}
we arrive at an effective lattice Hamiltonian
\begin{align}\label{eq:H}
H &= \sum_\vk \varepsilon_{\vk,b} b_\vk^\dag b_\vk + \sum_\vk \varepsilon_{\vk,f} c_\vk^\dag c_\vk\notag\\
&+\frac{U_b}{2N}\sum_{\{\vk_j\}} b_{\vk_1}^\dag b_{\vk_2}^\dag b_{\vk_3} b_{\vk_4} + \frac{U_{bf}}{N}\sum_{\{\vk_j\}} b_{\vk_1}^\dag b_{\vk_2} c^\dag_{\vk_3} c_{\vk_4}.
\end{align}
This procedure is justified when the optical potential is strong enough, typically $V_\alpha > E_\alpha^\text{rec}$, where 
\begin{align}
E_\alpha^\text{rec} = 2\hbar^2\pi^2/(\lambda^2m_\alpha)
\end{align}
is the atom recoil energy and $\lambda$ is the wavelength of the laser light. For later purposes, we define the optical trap depth 
$s_\alpha = V_\alpha/E_\alpha^\text{rec}$. In order to evaluate the critical superfluid velocity explicitly from the microscopic 
parameters of an experimental setup, we make use of following expression for the hopping and interaction parameters \cite{sengupta_pra_07}:
\begin{align}\label{eq:t}
t_\alpha &= \frac{2(E_\alpha^\text{rec}V_\alpha^3)^{1/4}}{\sqrt{\pi}\e{\sqrt{4V_\alpha/E_\alpha^\text{rec}}}},\notag\\
U_b &= \frac{\sqrt{32\pi}(E_b^\text{rec}V_b^3)^{1/4}a_b}{\lambda}.
\end{align}
For the fermion-boson interaction, one has
\begin{align}\label{eq:U}
U_{bf} = \frac{8\sqrt{\pi}(E_f^\text{rec} V_b^3V_f^3)^{1/4} (1+m_f/m_b) a_{bf}}{\lambda(\sqrt{V_b} + \sqrt{V_fE_b^\text{rec}/E_f^\text{rec}})^{3/2}}.
\end{align}
It is also useful to introduce the scattering lengths $\{a_b,a_{bf}\}$, which are related to the interaction parameters in Eq. (\ref{eq:Hint}) as follows:
\begin{align}
a_b = \frac{g_bm_b}{4\pi \hbar^2},\; a_{bf} = \frac{g_{bf}m_fm_b}{2\pi(m_f+m_b)\hbar^2}.
\end{align}
The onsite potentials in Eqs. (\ref{eq:t}) and (\ref{eq:U}) are obtained by relating them directly to the Wannier functions $\{\mathcal{U}(\vecr), \mathcal{V}(\vecr)\}$ used to approximate the wavefunctions in the lowest Bloch band. For instance, one has \cite{buchler_pra_04}:
\begin{align}
U_{bf} = g_{bf} \int \text{d}\vecr |\mathcal{U}(\vecr)|^2|\mathcal{V}(\vecr)|^2,
\end{align}
where we have defined
\begin{align}
\mathcal{U}(\vecr-\boldsymbol{R}) &= \frac{1}{N}\sum_\vk u_\vk(\vecr) \e{-\i\boldsymbol{R}\cdot\vk},\notag\\
\mathcal{V}(\vecr-\boldsymbol{R}) &= \frac{1}{N}\sum_\vk v_\vk(\vecr) \e{-\i\boldsymbol{R}\cdot\vk}.
\end{align}
The energy dispersions are dictated by the geometry of the optical lattice and are proportional to the nearest-neighbor matrix 
elements $t_\alpha$, whereas the summation over $\{\vk_j\}$ should be taken such that momentum is conserved in the scattering 
process.   
\par
The Hamiltonian Eq. (\ref{eq:H}) is now quadratic in the fermion-sector, which allows us to integrate out the fermions in the 
partition function by using a functional integral formulation. After doing so, one identifies an effective boson-boson interaction 
$U_{b}$ of the form \cite{buchler_prl_03}
\begin{align}
U_{b} &= U_b + U_{bf}^2\chi(T,\vq),\notag\\
\chi(T,\vq) &= \frac{1}{N} \sum_\vk \frac{F(\varepsilon_{\vk,f})-F(\varepsilon_{\vk+\vq,f})}{ \varepsilon_{\vk,f} - \varepsilon_{\vk+\vq,f} + \i\delta},\; \delta\to 0.
\end{align}
Here, $\chi(T,\vq)$ is the Lindhard function describing the fermionic polarization-bubble response, and $F(\varepsilon) = [1+\e{\beta(\varepsilon-\mu)}]^{-1}$, $\beta=(k_BT)^{-1}$ is the Fermi distribution function.
\par
To proceed analytically, we restrict ourselves to the weak-coupling regime and employ a Bogoliubov mean-field theory for 
superfluidity \cite{oosten_pra_01} to arrive at the bosonic quasiparticle excitation spectrum
\begin{align}
\mathcal{E}_{\vq,b} = \sqrt{\varepsilon_{\vq,b}\{\varepsilon_{\vq,b} + 2n_b[U_b + U_{bf}^2\chi(T,\vq)]\}}.
\end{align}
The phase-separation criterion (the point at which the bosonic excitation energies cease to be
real) thus reads
\begin{align}
U_b < -U_{bf}^2  ~ \lim_{q \to 0}\chi(T,\vq).
\end{align}
Phase-separation is triggered by the effective boson-boson interaction becoming attractive, which leads to a negative compressibility 
and an unstable homogeneous superfluid state \cite{abrikosov_book}. Note that the critical value of $U_{bf}$ where phase-separation 
sets in is independent of the sign of the interaction $U_{bf}$. As an example of how the phase-separation may be manifested, it was 
shown in Ref. \cite{molmer_prl_98} how the bosonic density in a Bose-Fermi mixture confined in a three-dimensional harmonic trap 
would be strongly enhanced in the center of the trap surrounded by a fermionic density-shell in the phase-separated regime. 
\par
Denoting the Fermi level by $\varepsilon_f$, one finds that
\begin{align}\label{eq:regular}
\chi(T,\vq\to 0) = \int \text{d}\varepsilon N(\varepsilon) \partial_\varepsilon F(\varepsilon) = -N(\varepsilon_f).
\end{align}
The effective boson-boson interaction then takes the form $U_\text{eff} = U_b - U_{bf}^2N(\varepsilon_f)$, and remains at a constant 
positive or negative value when varying the temperature in the regime $T\ll T_f$, where $T_f$ is the Fermi temperature. In a 
two-dimensional lattice structure, the energy bands feature saddle points at distinct wavevectors, thus giving rise to well-known 
van Hove singularities. In the vicinity of a van Hove singularity, the DOS is not a smooth function of the energy
and Eq. (\ref{eq:regular}) no longer holds. 
When the fermionic chemical potential is tuned to match the van Hove singularity, the Lindhard function diverges 
logarithmically as follows \cite{orth_arxiv_09}
\begin{align}
\chi(T\to 0,0) = -\chi_0\text{ln}\Big(\frac{\mathcal{C}t_f}{k_BT}\Big),
\end{align}
in the zero-temperature limit $T\to0$. Here, $\chi_0$ is a prefactor of dimension inverse energy, whereas $\mathcal{C}$ is a numerical prefactor. For a square lattice, one finds $\chi_0 = 1/(2\pi^2t_f)$ and $\mathcal{C} \simeq 18.08$, whereas for \eg a triangular lattice one would find $\chi_0 = 3/(4\pi^2t_f)$ and $\mathcal{C} = 9.04$ \cite{orth_arxiv_09}. It should be noted that we have considered the static limit $\i\varepsilon_n \to 0$ for the Lindhard function, where $\varepsilon_n$ is a bosonic Matsubara frequency. This approximation is valid for a scenario where the fermion 
response-time is much faster than the bosonic equivalent, which means that one can disregard retardation effects \cite{orth_arxiv_09}.

\begin{figure}[t!]
\centering
\resizebox{0.41\textwidth}{!}{
\includegraphics{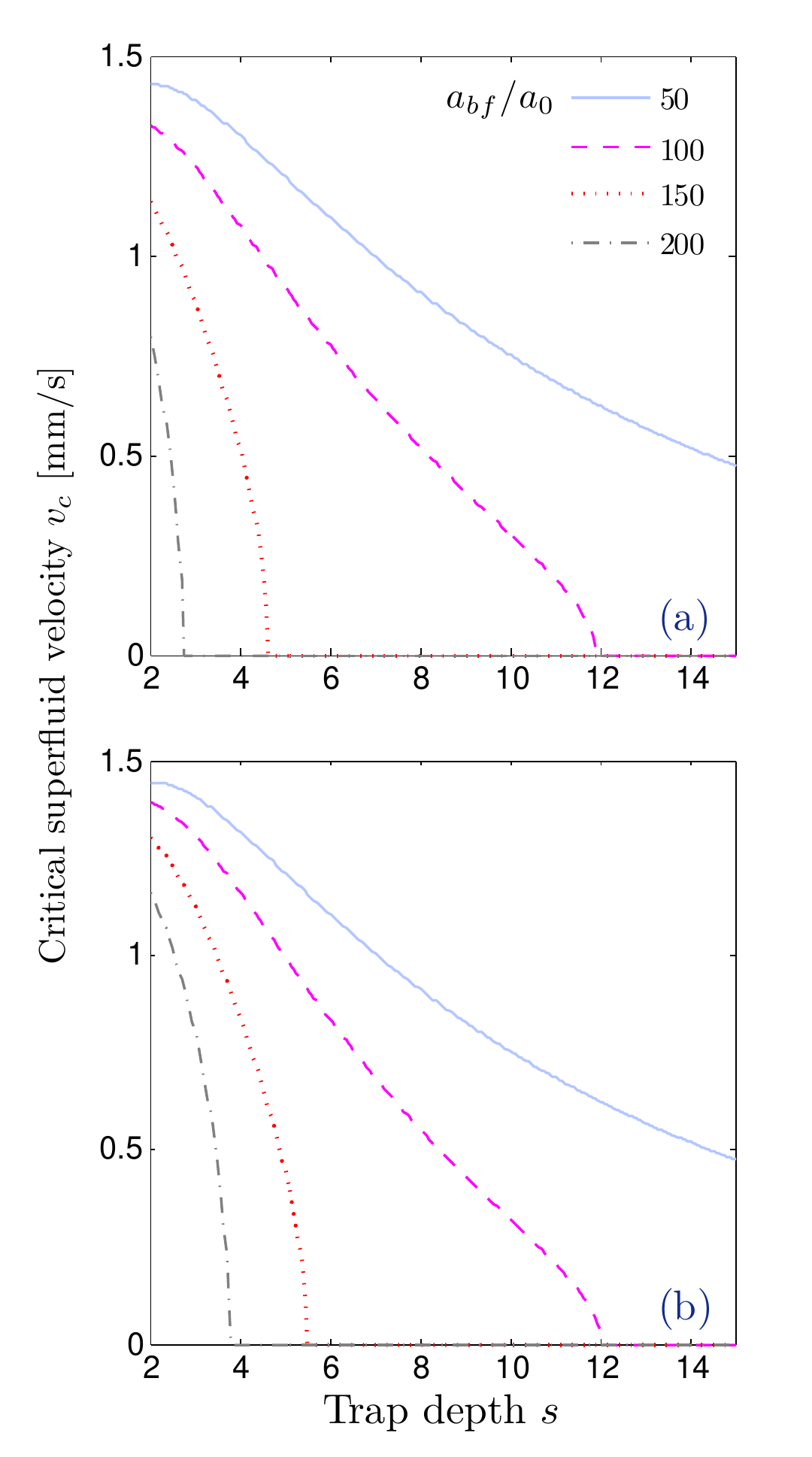}}
\caption{(Color online) Plot of the critical superfluid velocity $v_c$ for the bosons and its dependence on the trap depth. Here, we have set (a) $\mu/t_f = 0.5$ and (b) $\mu/t_f=3.5$. All other parameter values are specified in the main text.}
\label{fig:Velocity_svec}
\end{figure}

\begin{figure}[t!]
\centering
\resizebox{0.41\textwidth}{!}{
\includegraphics{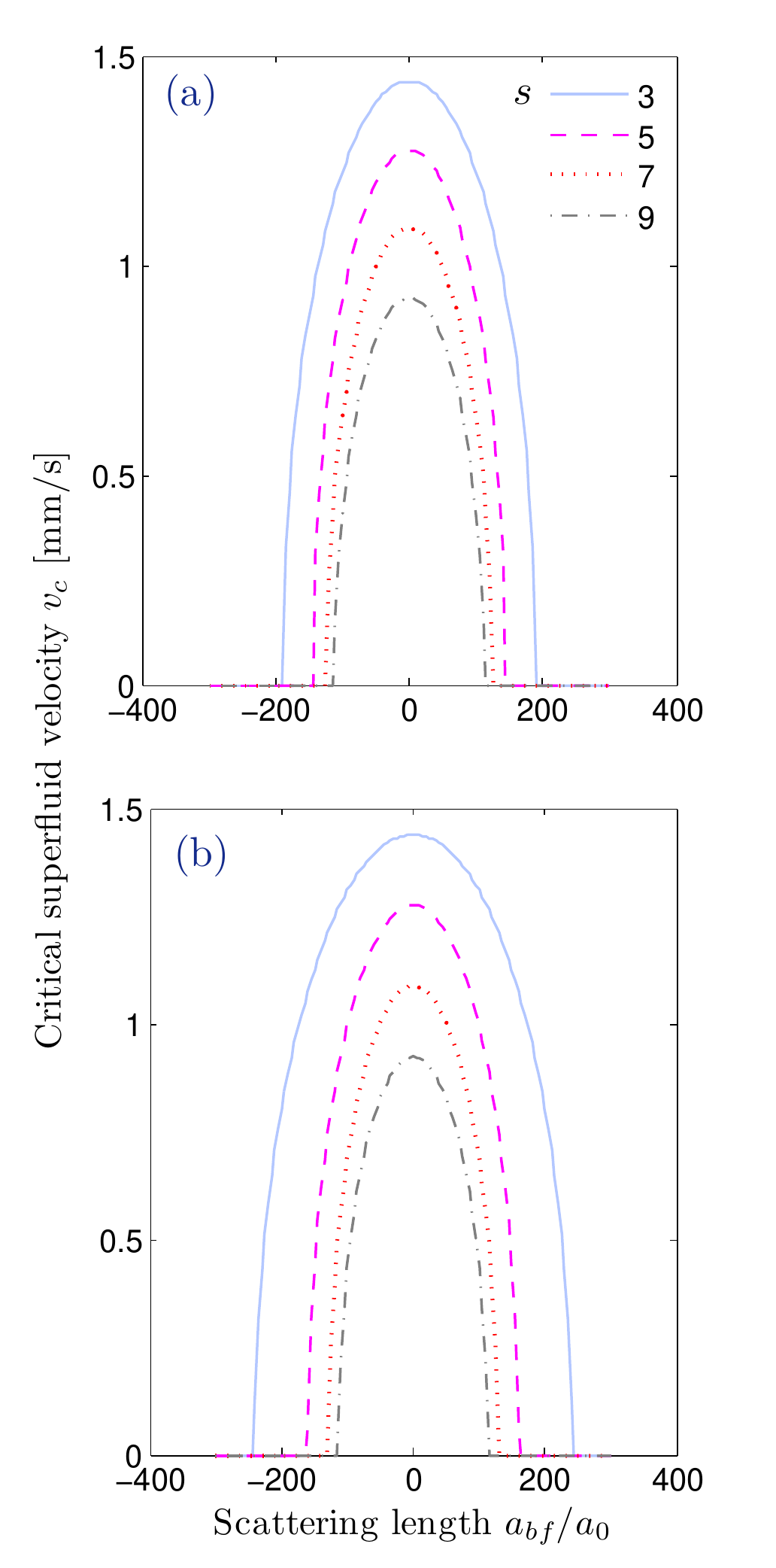}}
\caption{(Color online) Plot of the critical superfluid velocity $v_c$ for the bosons and its dependence on the fermion-boson interaction. Here, we have set (a) $\mu/t_f = 0.5$ and (b) $\mu/t_f=3.5$. All other parameter values are specified in the main text.}
\label{fig:Velocity_avec}
\end{figure}

\section{Results}\label{sec:results}

From now on, we will consider a simple square lattice for concreteness, which is the easiest setup to realize experimentally. In this case, the van Hove singularity is located at $\varepsilon=0$ and the DOS has a bandwidth of $W=8t_f$. We find that the energy dispersion in the long wavelength limit reads 
\begin{align}
\mathcal{E}_b(T,\vq\to 0) = \sqrt{4n_bt_ba^2[U_b + U_{bf}^2\chi(T,0)]}|\vq|.
\end{align}
The critical superfluid velocity $v_c$ is obtained in the standard way 
\begin{align}
v_c = \text{min}\Big(\frac{\mathcal{E}_b}{\hbar|\vq|}\Big),
\end{align}
leading to
\begin{align}
v_c &= \sqrt{4n_bt_ba^2[U_b + U_{bf}^2\chi(T,0)]}/\hbar.
\end{align}
Some properties of Bose-Fermi mixtures with a fermionic chemical potential tuned to the van Hove singularity were discussed in Refs. \cite{buchler_prl_03, buchler_pra_04}. Here, we will consider a situation of a non-zero chemical potential, thus moving away from half-filling. In order to model a realistic experiment, we will employ the following parameters for a $^{40}$K-$^{87}$Rb Bose-Fermi mixture \cite{roati_prl_02, kempen_prl_02}: $T_\text{BEC} = 100$ nK, $m_f = 6.64\times10^{-26}$ kg, $m_b = 1.44\times10^{-25}$ kg, $a_b \simeq 98a_0$. Here, $a_0 \simeq 52.9\times10^{-3}$ nm is the Bohr radius. In order to ensure equal lattice depths $s_\alpha\equiv s$ for the fermions and bosons, measured relative their respective recoil energies $E_\alpha^\text{rec}$, we fix $\lambda=755$ nm \cite{best_prl_09}. In general, the effective potentials seen by the fermions and bosons can be tuned by detuning the lattice wavelength relative the wavelengths $\lambda_{f(b)}$ of the fermions (bosons) according to \cite{illuminati_prl_04}
\begin{align}
\frac{V_b}{V_f} = \frac{\Gamma_f\lambda_f^4\Delta\lambda_b}{\Gamma_b\lambda_b^4\Delta\lambda_f}
\end{align}
where $\Gamma_\alpha$ is the natural linewidth, $\alpha=\{f,b\}$. We set the zero-temperature condensate fraction to $n_B(T=0) = 0.5$, and employ a mean-field approximation for its temperature-dependence $n_B(T)$. We will fix the temperature at $T/T_\text{BEC}=0.6$, which should be feasible to 
reach experimentally and still within the regime of validity for a mean-field approximation \cite{modugno_pra_03}. The remaining parameters that 
must be specified are the chemical potential and the boson-fermion scattering length. As shown in Fig. \ref{fig:DOS}, we will consider two fermion fillings characterized by $\mu/t_f=0.5$ (close to the van Hove singularity) and $\mu/t_f=3.5$ (close to the band edge), respectively. The 
boson-fermion scattering length $a_{bf}$ is tunable, as shown recently in Ref. \cite{best_prl_09}. By using a Feshbach resonance, scattering 
lengths in a range $\pm800a_0$ were reached. We shall therefore consider both positive and negative scattering lengths, reaching up to 
several hundreds of $a_0$. In order to evaluate the critical superfluid velocity, we employ a numerical solution of the expression:
\begin{widetext}
\begin{align}\label{eq:velocity}
v_c = \Bigg[4n_bt_ba^2\Bigg(U_b - \frac{\beta}{8\pi^2t_f} U_{bf}^2 \int^{4t_f}_{-4t_f} \int^{\pi/2}_0 \frac{\text{d}\varepsilon\text{d}\gamma}{\sqrt{[\cos^2\gamma + (\varepsilon\sin\gamma/4t_f)^2]\cosh[\beta(\varepsilon-\mu)]} } \Bigg)\Bigg]^{1/2}
\end{align}
\end{widetext}
As seen from Eq. (\ref{eq:velocity}), it becomes necessary to account properly for the finite temperature $T$ in order to describe the physical properties of Bose-Fermi mixtures, unlike the purely bosonic case.  
\par
In what follows, we will investigate how $v_c$ depends on the fermion-boson interaction parameter $a_{bf}$ and the trap depth $s$, using the set of experimentally realistic parameters described above. Consider first its dependence on the trap depth $s$, as shown in Fig. \ref{fig:Velocity_svec}. One of the main features is that $v_c$ exhibits a robustness towards the trap depth for relatively low values of $a_{bf}$. For high values of $s$, one would expect a transition into a Mott insulating state for commensurate fillings. When the interaction $a_{bf}$ becomes strong compared to the intrinsic bosonic repulsion $a_b$, any increase in trap depth $s$ is much more efficient in suppressing the critical velocity. We have distinguished between two fermionic fillings corresponding to $\mu/t_f=0.5$ and $\mu/t_f=3.5$ in Fig. \ref{fig:Velocity_svec}, in order to compare the cases with a chemical potential close to the van Hove singularity and close to the band edge, respectively. As seen, the difference is minor except at large values of the interaction $a_{bf}$, where $v_c$ is substantially reduced with $\sim50\%$ for a given trap depth, in addition to a much smaller critical trap depth 
$s$ where the superfluid velocity vanishes. 
\par
Next, we consider how the critical superfluid velocity $v_c$ is influenced by the fermion-boson interaction $a_{bf}$.  The result is shown in Fig. \ref{fig:Velocity_avec}. Since the critical velocity in Eq. (\ref{eq:velocity}) depends on $U_{bf}^2$, the sign of the interaction is irrelevant 
for the magnitude of $v_c$. Upon increasing the magnitude of the interaction $a_{bf}$, the superfluid velocity is strongly reduced and eventually vanishes, indicating a phase-separated regime. This may be understood physically bÿ noting that the contribution from the Lindhard function is 
negative in Eq. (\ref{eq:velocity}), meaning that the fermionic contribution to the induced boson-boson interaction is attractive. As the 
Bose-condensed phase is unstable towards attractive interactions, the critical velocity vanishes when the fermionic contribution eventually 
overtakes the intrinsic bosonic repulsion. In Ref. \cite{best_prl_09}, it was very recently experimentally demonstrated how the fermion-boson interaction $a_{bf}$ could be tuned in a well-defined manner over a wide range $\pm800a_0$ by exploiting a Feshbach resonance. It should 
therefore be experimentally viable to monitor the critical velocity $v_c$ as a function of the interaction $a_{bf}$ by using such 
techniques.
\begin{figure}[t!]
\centering
\resizebox{0.5\textwidth}{!}{
\includegraphics{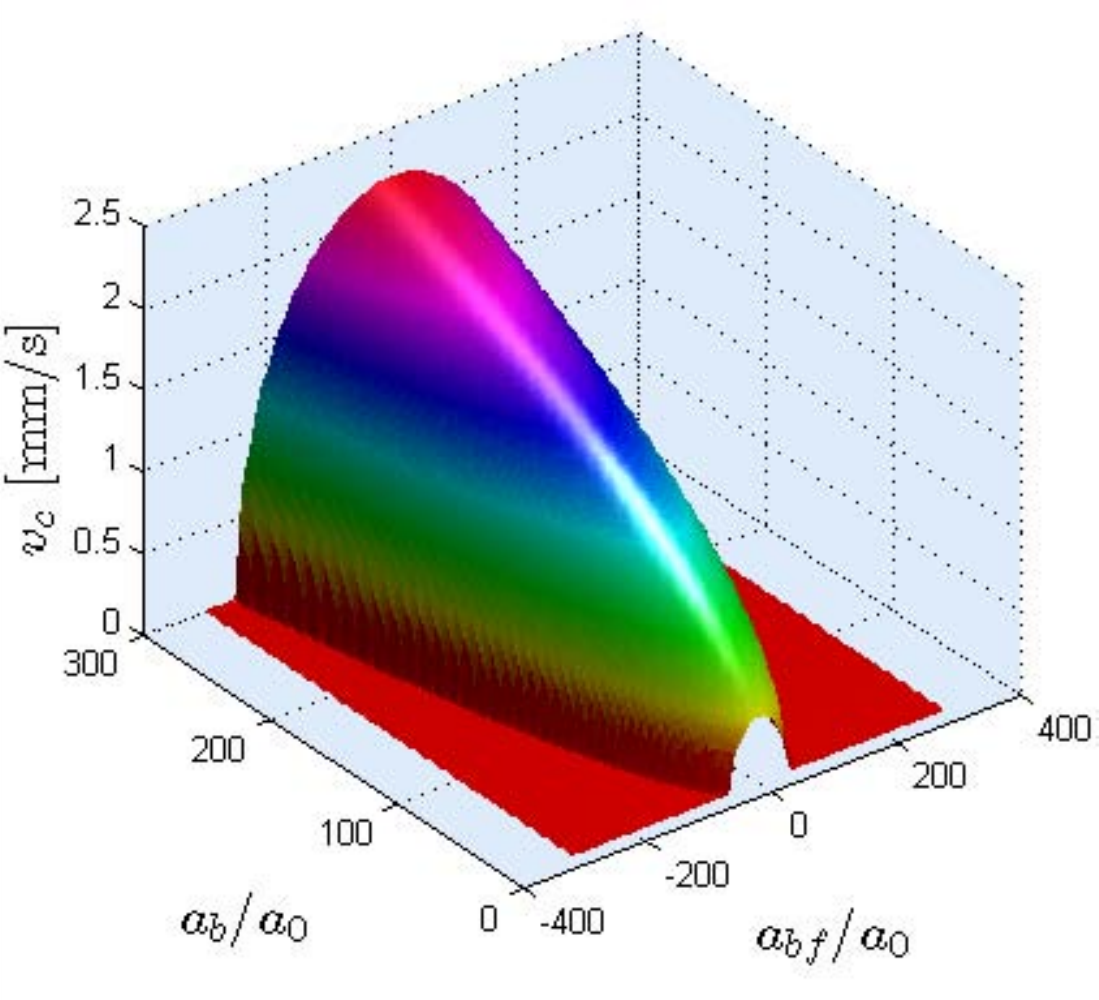}}
\caption{(Color online) Plot of the critical superfluid velocity as a function of the intrinsic boson-boson scattering length $a_b$ and the boson-fermion interaction $a_{bf}$. We have set $\mu/t_f=0.5$ and considered a trap depth $s=5$.}
\label{fig:Velocity_contour}
\end{figure}
\par
Finally, the mutual dependence on the intrinsic boson-boson interaction and the fermion-boson interaction is plotted in Fig. \ref{fig:Velocity_contour}, setting $\mu/t_f=0.5$ and $s=5$. While the critical velocity is suppressed with increasing $a_{bf}$, it is enhanced by increasing $a_b$. The reason for this is that the effective boson-boson interaction becomes more repulsive, in favor of the phase-coexistent state. It should nevertheless be emphasized that above a critical magnitude for the effective interaction $U_{b}$, a phase transition from superfluid to Mott insulator takes place. As shown in Ref. \cite{oosten_pra_01}, the present mean-field Bogoliubov approach does not capture this transition as it treats the interaction only in a weak-coupling regime. Therefore, the results reported here are obviously only valid inside the superfluid regime.

\section{Discussion}\label{sec:discussion}

\begin{figure}[t!]
\centering
\resizebox{0.5\textwidth}{!}{
\includegraphics{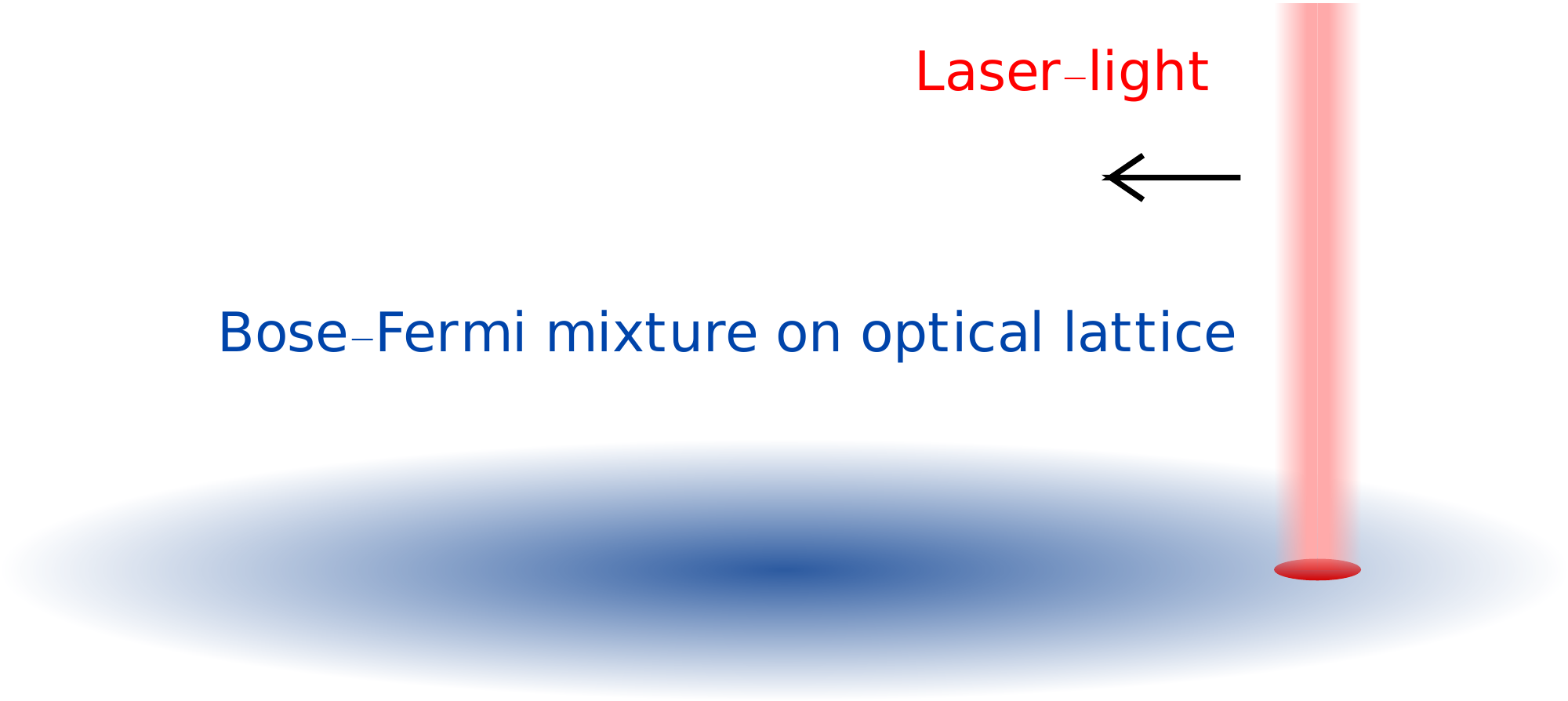}}
\caption{(Color online) Proposed experimental setup for probing the critical superfluid velocity in a Bose-Fermi mixture. A laser beam serves as a macroscopic object flowing through the condensate, thereby creating a moving boundary condition for the quasiparticle excitations.}
\label{fig:model}
\end{figure}

The experimental detection of a critical superfluid velocity requires measurements at temperatures well below $T_\text{BEC}$, thus in the nano-Kelvin regime. One possible route to probing the critical velocity was described in Ref. \cite{raman_prl_99}. There, dissipation in a Bose-Einstein condensed gas was monitored by means of moving a laser
beam through the condensate at different velocities (see Fig. \ref{fig:model}). The laser effectively plays the role of a massive macroscopic
object which creates a moving boundary condition. The main finding in Ref. \cite{raman_prl_99} was that strong heating was observed only above a critical velocity, and the laser was enabled to move back and forth through the condensate at a constant velocity by applying a triangular
waveform to an acousto-optic deflector.
\par
In the treatment of the critical superfluid velocity, it is implicitly assumed that the bosons are in the superfluid phase for the relevant parameter regime. In order to verify this, a full numerical self-consistent solution is required. Our main purpose here is to report on the magnitude of the critical velocity and its dependence on tunable parameters, using a set of realistic parameters employed in real experiments \cite{roati_prl_02, modugno_pra_03, best_prl_09} in which the bosons indeed were in the condensed state, which should justify our assumption. Finally, we note that the results we have obtained quantitatively agree with previous measurements for the critical velocity in Bose-Einstein condensates. In particular, $v_c\sim 1.6$ mm/s was reported in Ref. \cite{raman_prl_99}. 

\section{Summary}\label{sec:summary}

In conclusion, we have studied how the fermion-boson interaction for a Bose-Fermi mixture residing on an optical lattice modifies the effective interaction between the superfluid bosons. In particular, we have investigated how the phase-separation criteria is manifested through the critical superfluid velocity. Employing a set of microscopic parameters relevant to a $^{40}$K-$^{87}$Rb mixture \cite{roati_prl_02, modugno_pra_03, best_prl_09}, we report quantitative results for the magnitude of the superfluid velocity and its dependence on the trap depth, the boson-fermion interaction, and the fermionic filling fraction. All of these parameters can be tuned experimentally by means of the laser intensity and by exploiting Feshbach resonances. We find that the overall tendency of the boson-fermion interaction is to suppress $v_c$, and our quantitative results are of similar magnitude as previous measurements of the critical velocity in Bose-Einstein condensates, where $v_c \simeq 1.6$ mm/s was estimated \cite{raman_prl_99}. We have proposed an experimental setup for probing the critical superfluid velocity, which may serve as a direct tool to 
monitor a phase-separation scenario in a Bose-Fermi mixture.

\acknowledgments
J.L thanks I.B. Sperstad for helpful comments. This work was supported by the Research Council of Norway, Grants No. 158518/431 
and No. 158547/431 (NANOMAT), and Grant No. 167498/V30 (STORFORSK).

\end{document}